\documentclass[aps,pre,twocolumn]{revtex4}
\usepackage{graphicx}
\usepackage{psfrag}
\usepackage{amssymb}
\usepackage{amsmath}

\newcommand{\nnabla}{\mbox{\boldmath $\nabla$}}
\newcommand{\bc}{\begin{center}}
\newcommand{\ec}{\end{center}}
\newcommand{\ea}{\end{array}}
\newcommand{\ba}{\begin{array}}

\newcommand{\bee}{\begin{eqnarray}}
\newcommand{\eee}{\end{eqnarray}}
\newcommand{\x}{\mbox{\boldmath ${\zeta}$}}
\newcommand{\m}{\mbox{\boldmath ${\mu}$}}

\begin{document}
\title{
Lubrication approximation for micro-particles moving along parallel walls}
\author{Maria L. Ekiel-Je\.zewska}
\affiliation{Institute of Fundamental Technological Research,
    Polish Academy of Sciences, \'Swi\c etokrzyska 21, 00-049 Warsaw, Poland}
\author{Eligiusz Wajnryb}
 \affiliation{Institute of Fundamental Technological Research,
    Polish Academy of Sciences, \'Swi\c etokrzyska 21, 00-049 Warsaw, Poland}
\author{Jerzy B\l awzdziewicz}
\affiliation{Department of Mechanical Engineering, P.O. Box 20-8286, Yale University, New Haven,
CT 06520-8286, USA}
\author{Fran\c cois Feuillebois}
\affiliation{PMMH, UMR 7636 CNRS, ESPCI, 10 rue Vauquelin, 75231 Paris cedex 05, France}

\date{13.08.2008}

\begin{abstract}
Lubrication expressions for the friction coefficients of a spherical particle moving in a fluid 
between and along two parallel solid walls are explicitly evaluated in the low-Reynolds-number 
regime. They are used to determine lubrication expression for the particle free motion under 
an ambient Poiseuille flow. The range of validity and the accuracy of the lubrication 
approximation is determined by comparing with the corresponding results of the accurate 
multipole procedure. The results are applicable for thin, wide and long microchannels, or 
quasi-two-dimensional systems.
\end{abstract}

\maketitle
Microfluidic devices are commonly used for particle manipulation and separation, such as biological cell sorting, on-chip hydrodynamic chromatography~\cite{chromatography}, electrophoresis~\cite{electrokinetic}, and many other applications~\cite{dufresne,lobry,Lin,Cui:04,Diamant:05,Arauz:05,Pesch:00,pores1,pores2}. Particles often move along microchannels whose smallest dimension is comparable with the particles' size, and the other ones are much larger~\cite{Staben2,Staben3}. For such a quasi-two-dimensional system, it is of interest to theoretically determine the particle velocity, solving the Stokes equations for the fluid motion between two parallel infinite walls. 

Particle velocity can be evaluated numerically, by the boundary-integral method~\cite{Ganatos,Staben} or the multipole expansion~\cite{Poiseuille}. However, as pointed out in Ref.~\cite{Staben}, near-contact explicit asymptotic expressions would be useful, in analogy to the widely applied lubrication expressions for pairs of very close solid spherical particles of different sizes~\cite{JO}, including the limiting case of a sphere close to a single plane wall~\cite{Goldman,stewartson}. 

Therefore the goal of this paper is to derive lubrication expressions for a spherical particle moving along two close parallel solid walls, and check their accuracy.

\begin{figure}
[ht]
\includegraphics[height=3.5cm]{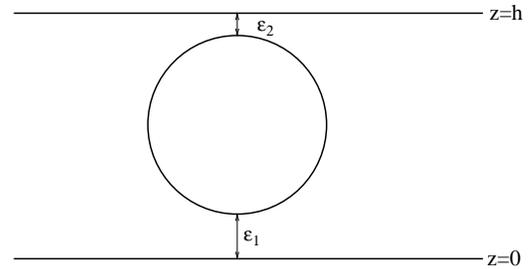}
\caption{The system and notation.}\label{1}
\end{figure}

A spherical particle of radius $a$ moving in a Stokes flow between
two parallel solid walls is considered, as illustrated in Fig.~\ref{1}. Distances are normalized by the particle radius $a$. The wall at $z=0$ is labeled 1, and the wall at $z=h>0$ is labeled 2. The  sizes of the gaps between the walls and the particle surface 
are small, $\epsilon_1 \ll 1$ and $\epsilon_2 \ll 1$. 

The fluid velocity ${\bf v}$ and pressure $p$ satisfy the Stokes equations~\cite{Kim,happel},
\bee
\eta \nnabla^{2} {\bf v} - \nnabla p = 0, \; \; \; \; \; \nnabla \cdot {\bf v} = 0,
\label{eq:stokes}
\eee
with the stick boundary conditions at the walls and the particle surface.

In the friction problem, 
there is no ambient flow and the particle translates along the channel with velocity $U_x \hat{\bf x}$ and rotates with angular velocity $\Omega_y \hat{\bf y}$. 
The goal it to find 
 the hydrodynamic force ${\cal F}_x\hat{\bf x}$ and torque 
${\cal C}_y\hat{\bf y}$ exerted by the particle on the fluid~\cite{Kim},
\bee
\left( 
\ba{c}
{\cal F}_x\\
{\cal C}_y/a
\ea
\right) &=& 8\pi \mu a \, \left[\x \cdot \left(
\ba{c}
U_x\\
a\Omega_y
\ea \right) \right].
\eee
Here $\x$ denotes the dimensionless 2$\times$2 friction matrix, 
\bee
\x &=& \left( \ba{ccc} 
\frac{3}{4} f^t_{xx}\; &  -c^t_{yx}
\vspace{0.1cm}\\
- c^t_{yx}\; & c^r_{yy}
\ea\right),\label{friction}
\eee
with the friction factors $f^t_{xx}$, $c^t_{yx}$, $c^r_{yy}$~\cite{ChFe,PSF,Accuracy} to be found. 

Free motion in Poiseuille flow
means 
that a spherical particle is entrained by the ambient Poiseuille flow,
\bee
{\bf v}_0/v_{\max}&=&4 z(z-h)/h^2\,\hat{\bf x}.
\eee
In the absence of external forces, the particle translates along $\hat{\bf x}$ and rotates along $\hat{\bf y}$. The translational and angular velocities, normalized by $v_{\max}$ and $v_{\max}/a$, respectively, have the form,
\bee
\left( \ba{c} u\\ \omega \ea\right)
&=& \m  
\cdot
\left( \ba{c}
\frac{3}{4}f^p_{xx}\\ 
c^p_{yx}
\ea\right),\label{main}
\eee
with the mobility matrix obtained from Eq.~\eqref{friction},
\bee
\m=\x^{-1},\label{mu}
\eee
and the dimensionless friction factors $f^p_{xx}$, $c^p_{yx}$ determined by the force ${\cal F}^p_x\hat{\bf x}$ and torque ${\cal C}^p_y\hat{\bf y}$ exerted by the Poiseuille flow on immobile sphere~\cite{Accuracy} in the following way,
\bee
\left(
\ba{c}
{\cal F}^p_x\\
{\cal C}^p_y/a
\ea
\right) 
&=&
8\pi\mu a v_{\max} \left( \ba{c} 
\frac{3}{4}f^p_{xx}\\ 
c^p_{yx}
\ea\right).
\eee

In this work, the functional dependence of the friction factors and the particle's free-motion velocities on $\epsilon_1$ 
and $\epsilon_2$ has been determined numerically. 
The accurate theoretical 
method~\cite{2 walls JFM,2 walls Physica} of solving the Stokes equations, based on the  
multipole expansion~\cite{CFHWB}, has been applied. For a system of particles between two parallel walls, 
this method involves expanding the fluid velocity field into spherical and Cartesian 
fundamental sets of Stokes flows. The spherical set is used to describe the interaction 
of the fluid with the particles and the Cartesian set to describe the interaction
with the walls. At the core of the method are transformation relations between
the spherical and Cartesian fundamental sets. The transformation formulas are used to 
derive a system of linear equations for the force multipoles induced on the
particle surfaces. The coefficients in these equations are given in terms of lateral
Fourier integrals corresponding to the directions parallel to the walls. 
These equations are truncated at an multipole order $L$ \cite{CFHWB} and solved numerically, 
using the algorithm and the FORTRAN code described in 
Refs.~\cite{2 walls JFM,2 walls Physica,Poiseuille} and available at \cite{code}. 

To derive the lubrication approximation, the 
multipole algorithm described above has been first used to compute the friction matrix $\x$, given by Eq. (\ref{friction}). 
The friction matrix has been evaluated as the sum of two terms,
\bee
 \x(\epsilon_1,\epsilon_2) &=&  \left( \sum_{i=1}^2 \x_{i}(\epsilon_i) -\x_0, \right)+\x_{12}(\epsilon_1,\epsilon_2).\label{delta}
\eee
The first one, {\it the single-wall superposition}, is the sum of the one-particle friction matrices in a fluid bounded by the wall $i$ only,
\bee
\x_{i} &=& 
\left( \ba{ccc} 
\frac{3}{4} f^t_{i,xx}\; & \; -c^t_{i,yx}
\vspace{0.3cm}\\
- c^t_{i,yx}\; &\; c^r_{i,yy}
\ea\right),
\eee
minus the one-particle friction matrix in unbounded fluid,
\bee
\x_0 = 
\left(\ba{cc}
\frac{3}{4}&0\\
0&1
\ea \right).
\eee

The second one, $\x_{12}$, is the two-wall contribution, 
\bee
\x_{12}&=& \left( \ba{cc} \frac{3}{4} f^{t}_{12,xx} & -c^{t}_{12,yx}\\
-c^{t}_{12,yx}& c^{r}_{12,yy}
\ea
\right).\label{dec}
\eee
The single-wall near-contact expressions for $\x_{i}(\epsilon_i)$ are known~\cite{Goldman,stewartson,ChFe,CichockiJones},
\bee
f^{t}_{i,xx}(\epsilon_i)&=& -\frac{8}{15}\ln \epsilon_i + 0.95429-\frac{64}{375}\epsilon_i \ln \epsilon_i + 0.42945\,\epsilon_i,\nonumber\\\\
c^{t}_{i,yx}(\epsilon_i) &=& (-1)^{i+1} \left(-\frac{1}{10}\ln \epsilon_i - 0.19295 -\frac{43}{250}\epsilon_i \ln \epsilon_i \right.\nonumber\\
 &+& \left. 0.10058\,\epsilon_i\right),
\\
c^{r}_{i,yy}(\epsilon_i)&=&-\frac{2}{5}\ln \epsilon_i + 0.37089 -\frac{66}{125}\epsilon_i \ln \epsilon_i +0.34008\,\epsilon_i.\nonumber\\\label{lub1w}
\eee
The task is to derive near-contact approximation of 
the remaining two-wall contribution $\x_{12}$, regular in $\epsilon_i$.  
The components of $\x_{12}(\epsilon_1,\epsilon_2)$ are linear functions of $\epsilon_1+\epsilon_2$ or $\epsilon_1-\epsilon_2$, depending on symmetry.
 To determine them explicitly, values at the contact, $\x_{12}(0,0)$, have been first evaluated for all the subsequent multipole orders $L \le 404$, and extrapolated to $L= \infty$, see Refs.~\cite{CFHWB,Accuracy} for the basic concepts. Next, 
the multipole numerical codes with a very high  multipole order $402\le L\le 404$ have been applied to evaluate $\x_{12}(\epsilon_1,\epsilon_2)$ as a function of $0\le\epsilon_1\le 0.02$ for $\epsilon_2=0$.
 The following linear relations have been found,
\bee
f^{t}_{12,xx}(\epsilon_1,\epsilon_2) &\approx& 0.52805 -0.64 \,(\epsilon_1 +\epsilon_2),\label{twof}\\
c^{t}_{12,yx}(\epsilon_1,\epsilon_2)&\approx&0.175\,(\epsilon_1-\epsilon_2),\\
c^{r}_{12,yy}(\epsilon_1,\epsilon_2) &\approx& -0.06643 +0.08 \,(\epsilon_1+\epsilon_2).
\eee
with the accuracy $0.52805\pm0.0001$, $-0.64\pm0.01$, $0.175\pm0.005$, $-0.06643\pm 0.00002$ and $0.08\pm0.01$. 

Finally, the two-wall lubrication asymptotics of the total friction matrix~\eqref{friction} 
is the following,

\bee
f^{t}_{xx} \!\!\!&\approx& \!\!1.4366 -\frac{8}{15}\ln (\epsilon_1 \epsilon_2)
-\frac{64}{375}(\epsilon_1 \ln \epsilon_1+\epsilon_2 \ln \epsilon_2)\nonumber\\
&-&0.21 \,(\epsilon_1 +\epsilon_2),\label{lub1}\\
c^{t}_{yx}\!\!\!&\approx&\!\!-\frac{1}{10}\ln \frac{\epsilon_1}{\epsilon_2}
-\frac{43}{250}(\epsilon_1 \ln \epsilon_1\!-\!\epsilon_2 \ln \epsilon_2)+0.276 \,(\epsilon_1\! -\!\epsilon_2),\nonumber \\\label{lub2}\\
c^{r}_{yy}\!\!\!&\approx&\!\!-0.32465 -\frac{2}{5}\ln (\epsilon_1 \epsilon_2)-\frac{66}{125}(\epsilon_1 \ln \epsilon_1+\epsilon_2 \ln \epsilon_2) \nonumber \\
\!\!\!&+&\!\!0.42 \,(\epsilon_1\! +\!\epsilon_2).\label{lub3}
\eee

For applications, the most important is the translational coefficient $f^t_{xx}$. For the channel widths $h=2.1-2.5$,  the logarithmic term, $-\frac{8}{15}\log(\epsilon_1\epsilon_2)$, give only 68.5-50.5\% of the total lubrication expression~\eqref{lub1}.
The constant term, 1.4366, gives the additional 30.8-49.1\% (the other terms are practically negigible). Note that to evaluate this constant precisely, it is essential to compute the two-wall contribution \eqref{twof} at the contact, which is as large as 0.52805, i.e. 10.0-7.1\% of the total lubrication expression~\eqref{lub1}.

Derivation of the lubrication formulas for the free motion in Poiseuille flow will now be discussed. 
First, the near-contact approximations to the force $f^p_{xx}(\epsilon_1,\epsilon_2)$ and torque $c^p_{yx}(\epsilon_1,\epsilon_2)$ exerted by the ambient Poiseuille flow on the motionless particle are constructed, using the accurate numerical results.
In the absence of relative motion, there is no non-analytic terms. For $\epsilon_1=\epsilon_2=0$, the force is evaluated 
for subsequent multipole orders $L\le 404$. The linear relation of the force on $L^{-3}$, given in Ref.~\cite{Accuracy}, is now used to extrapolate the results until $L=\infty$, and to determine $f^p_{xx}(0,0)$.
The next term is linear in $(\epsilon_1+\epsilon_2)$. To determine the coefficient, the function  $f^p_{xx}(\epsilon_1,0)$ is evaluated 
with $402\le L\le 404$, and the linear dependence is found in the range $0\le \epsilon_1\le 0.005$. 
Similar procedure is repeated for $c^p_{yx}(\epsilon_1,0)$. In this case, however, quadratic terms, proportional to $(\epsilon_1^2-\epsilon_2^2)$, are needed in addition to those linear in $(\epsilon_1-\epsilon_2)$, because,  owing to symmetry, $c^p_{yx}(0,0)\!=\!0$. The near-contact asymptotics reads,
\bee
f^p_{xx}&\approx& 2.67817756 -1.1035 (\epsilon_1+\epsilon_2),\label{flub}\\
c^p_{yx}&\approx&-0.541(\epsilon_1-\epsilon_2)+0.5(\epsilon_1^2-\epsilon_2^2).\label{club}
\eee

Finally, lubrication approximation of the free motion velocities is obtained from Eqs.~\eqref{main}-\eqref{mu}, with the use of the near-contact expressions~\eqref{lub1}-\eqref{club}.

The range of validity of the above lubrication approximation has been determined by comparison with the numerical results. 
First, motion 
in a narrow channel of a fixed width, 
$
h=\epsilon_1+\epsilon_2+2,
$ with $h=2.1,\;2.2,\;2.3,\;2.4,\;2.5$,  has been examined. 
The errors of the near-contact expressions are listed in Table~\ref{tab}.
\begin{table}[ht]
\caption{
The maximal absolute and relative errors of 
the lubrication approximation for different channel widths $h$.}\label{tab}
\begin{tabular}{|c|l|l|l|l|l|}
\hline
$h$&2.1&2.2&2.3&2.4&2.5\\
\hline
\hline
$\delta f^{t}_{xx}$&0.004&0.02&0.04&0.06&0.09\\
\hline
$\delta c^{t}_{yx}$&0.0003&0.002&0.005&0.008&0.01\\
\hline
$\delta c^{r}_{yy}$&0.007&0.02&0.045&0.07&0.10\\
\hline
$\delta f^p_{xx}$&0.006&0.02&0.05&0.085&0.13\\
\hline
$\delta c^p_{yx}$&0.0003&0.002&0.007&0.015&0.03\\
\hline
$\delta u$     &0.0008& 0.003&0.007& 0.013  &0.02\\
\hline
$\delta \omega$&0.00007&0.0007&0.003& 0.007 &0.01\\
\hline
$\delta f^{t}_{xx}/f^{t}_{xx}$&0.09\%&0.4\%&1\%&2\%&3\%\\
\hline
$\delta c^{t}_{yx}/f^{t}_{xx}$&0.005\%&0.04\%&0.1\%&0.2\%&0.3\%\\
\hline
$\delta c^{r}_{yy}/c^{r}_{yy}$&0.2\%&0.9\%&2\%&3\%&5\%\\
\hline
$\delta f^p_{xx}/f^p_{xx}$&0.2\%&1\%&2\%&4\%&5.5\%\\
\hline
$\delta c^p_{yx}/f^p_{xx}$&0.01\%&0.1\%&0.3\%&0.7\%&1.3\%\\
\hline
$\delta u/u$          &0.15\%&0.5\%&1\%& 2\%   &3\%\\
\hline
$\delta \omega/u$&0.01\%& 0.1\%&0.5\%&  1\%   &2\%\\
\hline
\end{tabular}
\end{table}
The absolute errors of $f^t_{xx}$,  $f^p_{xx}$ and $u$ are the largest in the center of the channel, of $c^t_{yx}$ and $c^r_{yy}$ -- at the contact, and of $c^p_{xx}$ and $\omega$ -- in between.  
Note that the physical relative errors are   $\delta c^{t}_{yx}/f^{t}_{xx}$,  $\delta c^{p}_{yx}/f^{p}_{xx}$ and $\delta \omega/u$ rather than $\delta c^{t}_{yx}/c^{t}_{yx}$, $\delta c^{p}_{yx}/c^{p}_{yx}$ and  $\delta \omega/\omega$, respectively. 

\begin{figure}[ht]
\includegraphics[width=8.6cm]{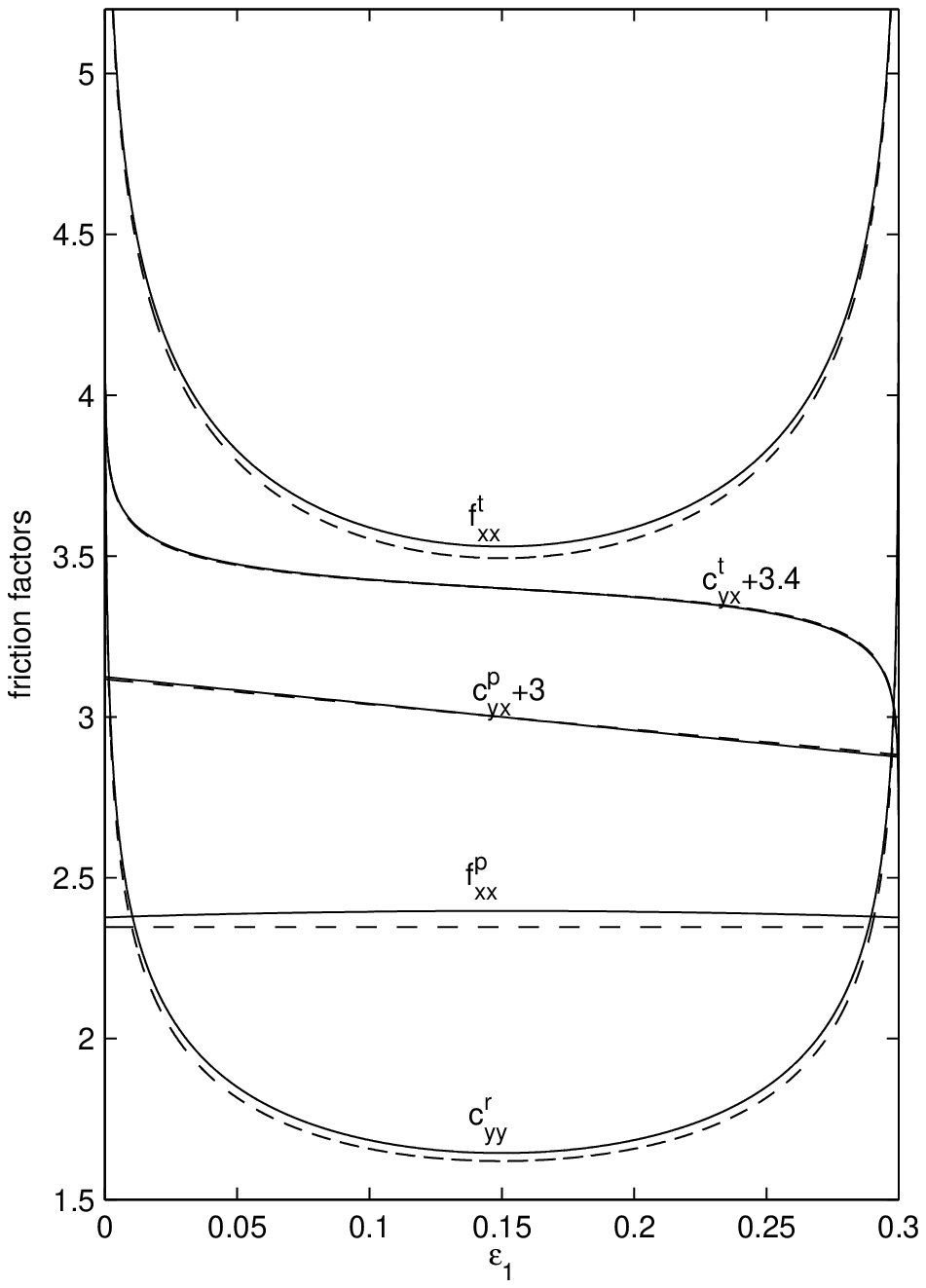}\\
\includegraphics[width=8.6cm]{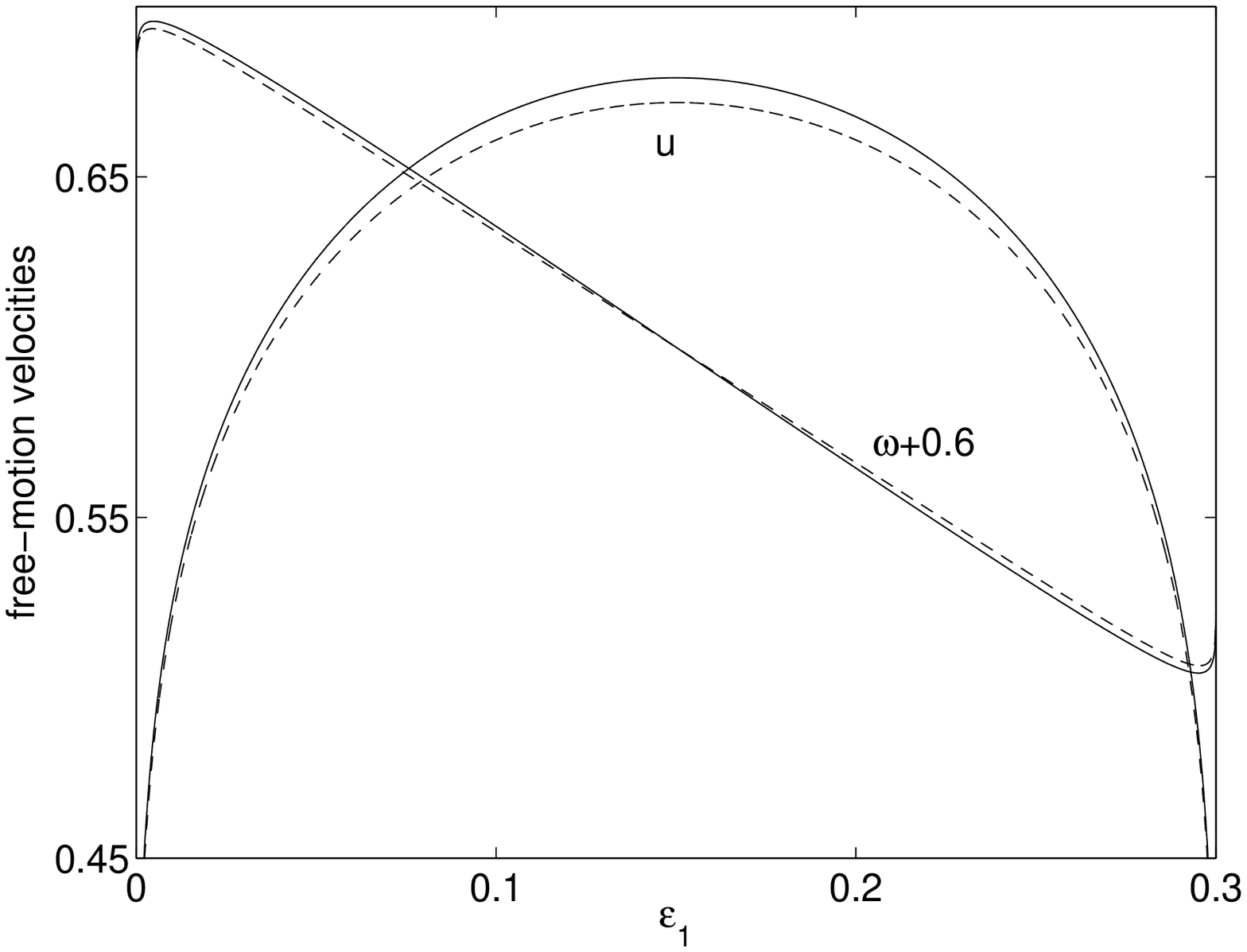}
\vspace{-0.5cm}
\caption{Friction coefficients (top) and translational and angular velocities of a freely moving sphere (bottom) in the channel of width $h=2.3$. 
Solid lines: accurate;  dashed line: near-contact.}\label{fric}
\end{figure}

In Fig.~\ref{fric}, the accurate friction factors 
and free motion velocities are plotted as functions of $\epsilon_1$, together with their lubrication approximation, given by Eqs.~\eqref{lub1}-\eqref{club} and~\eqref{main}-\eqref{mu}. Even for a relatively wide channel with $h=2.3$, the errors of the near-contact  physical expressions do not exceed~2\%.

\begin{figure}[ht]
\includegraphics[width=8.6cm]{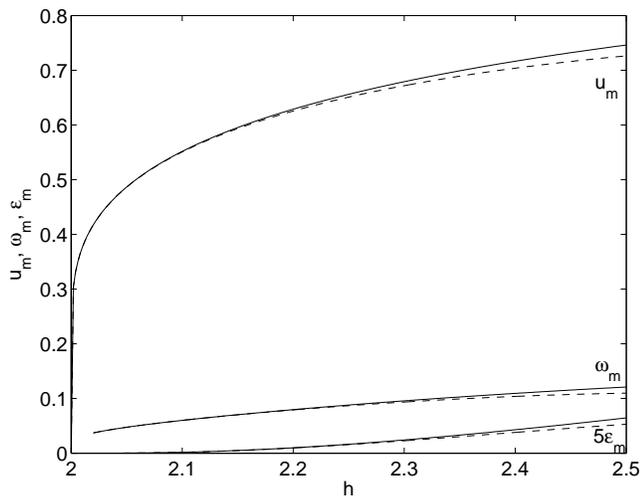}
\caption{Maximal translational and angular velocities of a freely moving sphere, $u_m$ and $\omega_m$, and the position of the maximal rotation, $\epsilon_m$, 
versus the channel 
width $h$. Solid lines: accurate;  dashed line: near-contact.}\label{velm2}
\end{figure}

In Fig.~\ref{velm2}, the maximal free motion velocities, $u_m\!=\!\underset{\epsilon_1}{\max}\, u(\epsilon_1)$ and 
$\omega_m\!=\!\underset{\epsilon_1}{\max}\,\omega(\epsilon_1)$ as well as the position of the largest rotation, $\epsilon_m\!\!:\,\omega(\epsilon_m)\!=\!\omega_m$, are plotted versus the channel width
$h$. Note that $u$ is the largest at the channel center, 
and $\omega$ -- very close to the wall.
For $h\lesssim 2.5$, the lubrication approximation of the maximal particle-velocities is still reasonably accurate, with $\delta u_m/u_m$, $\delta \omega_m/u_m< 3\%$.

\newpage

\begin{figure}[ht]
\includegraphics[width=8.6cm]{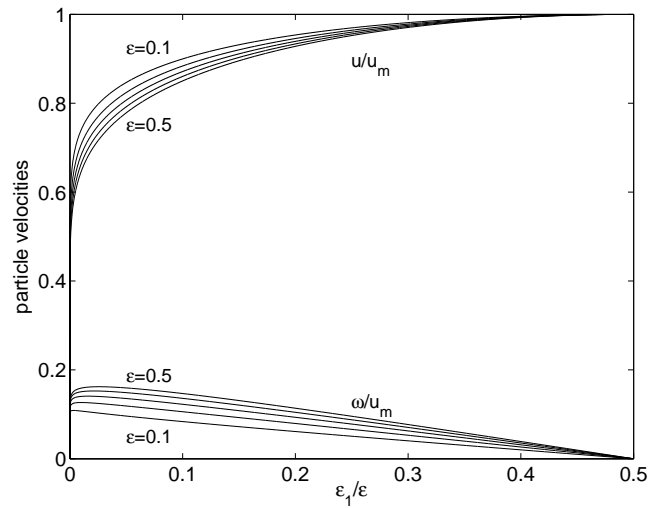}
\caption{Translational and rotational particle-velocity profiles for channels widths $h\!=\!2\!+\!\epsilon$, with
$\epsilon=0.1, 0.2, 0.3, 0.4$ and $0.5$ (top-down for $u/u_m$ and down-top for $\omega/u_m$).}\label{po}
\end{figure}

Finally,  velocity profiles of a freely moving 
sphere are compared for different channel widths. In the computation, $L=40$, with the $10^{-4}$ absolute accuracy.
Both translational and rotational velocities of the sphere, $u$ and $\omega$, are normalized by $u_m$, and the gap $\epsilon_1$ between the sphere surface and the lower wall -- by the sum of both gaps, $\epsilon=\epsilon_1+\epsilon_2$. 
The results are plotted in Fig.~\ref{po}.
The narrower channel, the closer to the plug flow translational velocity is observed, and the smaller ratio of $\omega/u$. 

Concluding, in this paper the near-contact approximation for a solid-sphere motion along parallel hard walls has been derived, with at least 2\% precision for the distance between the walls up to 2.3 radii. 

\vspace{-0.7cm}
\acknowledgements
\vspace{-0.5cm}
This work was supported in part by the Polish Ministry of Science and High Education, grant N501 020 32/1994.

\end{document}